\documentclass[preprint,aps,floats,amssymb,showpacs]{revtex4}
\usepackage{epsfig}

\newcommand{\be}{\begin{equation}}
\newcommand{\ee}{\end{equation}}
\newcommand{\bea}{\begin{eqnarray}}
\newcommand{\eea}{\end{eqnarray}}
\newcommand{\bml}{\begin{mathletters}}
\newcommand{\eml}{\end{mathletters}}

\begin{document}




\title{Fullerenic Solitons}
\renewcommand{\thefootnote}{\fnsymbol{footnote}}
\author{Yves Brihaye\footnote{Yves.Brihaye@umh.ac.be}}
\affiliation{Facult\'e des Sciences, Universit\'e de Mons-Hainaut,
7000 Mons, Belgium}
\author{Betti Hartmann\footnote{b.hartmann@iu-bremen.de}}
\affiliation{School of Engineering and Sciences, International University
Bremen (IUB), 28725 Bremen, Germany}

\date{\today}
\setlength{\footnotesep}{0.5\footnotesep}

\begin{abstract}
We study a modified non-linear Schr\"odinger equation on a 2 dimensional
sphere with radius $R$ aiming to describe electron-phonon
interactions on fullerenes and fullerides. These electron-phonon
interactions are known to be important for the explanation
of the high transition temperature of superconducting fullerides.
Like in the $R\rightarrow \infty$ limit, we are able to construct
non-spinning as well as spinning solutions which are characterised by the
number of nodes of the wave function. These solutions are closely related
to the spherical harmonic functions. For small $R$, we discover specific 
branches of the solutions. Some of the branches survive in the 
$R\rightarrow\infty$ limit and the solutions obtained on the plane 
($R=\infty$) are recovered. 

\end{abstract}
\maketitle
\renewcommand{\thefootnote}{\arabic{footnote}}
\section{Introduction}
Carbon exists in several forms in nature. One is the so-called
fullerene which was discovered for the first time in 1985 \cite{cks}.
Fullerenes are carbon-cage molecules in which a big number $n$ of carbon (C) atoms
are bonded in a nearly spherically symmetric configuration.
These C$_n$ configurations typically have diameters of $d\approx 7-15$ \AA \  and 
consist of 12 pentagons and $(\frac{n}{2}-10)$ hexagons, where $n\geq 24$
has to be even.
The most prominent example is the C$_{60}$ fullerene which also
is called a ``buckminster'' fullerene.  It consists
of 20 hexagons and 12 pentagons and has a diameter of $d_{60}\approx 7$ \AA.
In  these fullerenes, three of the valence electrons of 
the C-atom are used to form the bounds with
the neighbouring C-atoms, while the remaining ``free'' valence electron
can ``hop'' along the $n$ positions in the fullerene.

Due to unoccupied electron orbitals, it is easy to reduce fullerenes, i.e.
put extra electrons on them. It was found that if alkali metal atoms-which donate
one electron each- such
as rubidium (Rb), potassium (K), caesium (Cs) or sodium (Na) are put onto
a C$_{60}$ fullerene, this leads to a metallic or even superconducting behaviour
\cite{haddon}. The transition temperature $T_c$ of these alkali-doped
fullerenes, also called {\it fullerides}, is rather high for superconductors,
e.g. $T_c=33$ K for a RbCs$_{2}$C$_{60}$ \cite{tanigaki}.
The superconductivity can be explained by phonon-electron interactions
in the fullerides \cite{varma}. 
In \cite{harigaya} it was found that if a C$_{60}$ or a C$_{70}$ molecule
is doped with one or two excess electrons, the additional
charges accumulate nearly along an equatorial line of the molecule.
The distortion in the lattice and the electron distribution was found
to be polaron-like.

Recently, a modified non-linear Schr\"odinger equation has been studied in the
context of fullerene-related structures, so-called nanotubes \cite{hz,bhz}.
These objects are graphite-like sheets roled to form a cylinder.
The carbon atoms are bonded in a hexagonal lattice on this sheet.
The existence of solitons which are created through the interaction
of the ``excitation'' described by a complex scalar field $\psi$
and the lattice vibrations was discussed. The creation of these type of
solitons was introduced by Davydov in the 1970s \cite{davy} to explain
the dispersion free energy transport in biopolymers. $\psi$ then
is an amide I-vibration. Here, however, we think of
$\psi$ as the wave function of an ``excess'' electron interacting with
the lattice via electron-phonon interactions. The electron could e.g.
be thought of as the valence electron of an alkali metal atom with which 
the fullerene is doped. 
The full discrete system of equations (which results
from a Fr\"ohlich type Hamiltonian)
can be approximated by the above mentioned modified non-linear Schr\"odinger
equation on the 2-dimensional plane.

In this paper, we study the modified non-linear Schr\"odinger equation
for the hexagonal lattice on a 2-dimensional sphere, thus aiming to
describe the electron-phonon interactions in fullerenes and fullerides.
These interactions are, as mentioned previously, of particular interest for
the superconductivity of the fullerides.
Our model is, of course, an approximation since
fullerenes consist of hexagons and pentagons. However, since a) there are
more hexagons in the C$_{60}$ and b) the coefficients appearing in the non-linear
Schr\"odinger equation don't depend strongly on the actual form of the lattice,
we believe that this is a good approximation.

Finally, let us mention that  studying solitons on a sphere can
appear to be useful also from a mathematical point of view, as 
was stressed
e.g. in \cite{dein}. In the present case this appears to be true
as well. In particular, we notice a strong relationship between most of
the various
branches of solutions and the spherical harmonics.

Our paper is organised as follows: in Section II, we present
the modified non-linear Schr\"odinger equation on the sphere, in Section III,
we discuss non-spinning solutions and in Section IV spinning generalisations.
We give our conclusions in Section V.
 
\section{Modified Non-linear Schr\"odinger equation on the sphere}
We consider the following modified non-linear Schr\"odinger equation in 2 dimensions
\cite{hz,bepz}, which in dimensionless variables reads:
\begin{equation}
\label{cnls}
i\frac{\partial \psi}{\partial t}+\Delta \psi +ag\psi\left(\vert\psi\vert^2+
b\Delta \vert\psi\vert^2\right)=0
\end{equation}
where 
$a$, $b$ are constants with $a=2$, $b=\frac{1}{12}$ for the square lattice
studied in
\cite{bepz} and $a=4$,
$b=\frac{1}{8}$ for the hexagonal lattice studied in \cite{hz}.
The additional term proportional to $b$ results from both the discreteness of the
lattice and the interaction between the lattice and the complex scalar field
$\psi$ (``phonon-electron'' interaction).
 $g$ is the (``phonon-electron'') coupling constant of
the system
and determines the ``strength'' of the non-linear character of the equation.
Especially, as can be seen from the derivation of the modified non-linear
Schr\"odinger equation from the discrete equations \cite{hz}, the mass
of the carbon atom is encoded into $g$.

As 
stated in the introduction, we will study the above equation on a 2-dimensional
sphere
with radius $R$. The Laplacian operator then reads:
\begin{equation}
\Delta=\frac{1}{R^2}\left(\frac{\partial^2}{\partial\theta^2}+
\cot\theta\frac{\partial}{\partial\theta}+
\frac{1}{\sin^2\theta}\frac{\partial^2}{\partial\varphi^2}\right)  \ .
\end{equation}
As usual, the angles are such that $\theta$ $\epsilon$ $[0:\pi]$
and $\varphi$ $\epsilon$ $[0:2\pi]$ and $\psi=\psi(t,\theta,\varphi)$
is a complex valued scalar field.
Since we are mainly concerned
with fullerenes here which consist (next to pentagons) out of hexagons, we use
in analogy to \cite{hz} $a=4$ and $b=\frac{1}{8}$. The solutions of (\ref{cnls}) can be characterised by
their norm $\eta$:
\begin{equation}
\eta^2=R^2\int |\psi|^2 \sin\theta d\theta d\varphi
 \end{equation}
 as well as by their energy:
 \begin{equation}
 E=R^2\int \left( |\vec{\nabla}\psi|^2-\frac{a}{2}g |\psi|^4 + 
2bg(\vec\nabla |\psi|^2)^2\right) \sin\theta d\theta d\varphi \ .
 \end{equation}
 The coupling constant $g$ can be absorbed into the scalar field $\psi$
 by rescaling it as follows: $\psi\rightarrow\psi/\sqrt{g}$.
 In our  numerical simulations, we will set $g=1$. 
This in turn means that we drop 
 the normalisation
 constraint ($\eta=1$) for the wave function $\psi$. However, since we 
present
 explicitely the norm of our numerical solutions, the normalized 
 wave function and the corresponding value of $g$ can be easily computed.
Especially, if a solution ceases to exist for a specific value of the
norm, this means that the soliton solution exists only above a high enough value
of the phonon-electron coupling.
 \section{Non-spinning solutions}
 To construct explicit solutions, 
we use the following axially
symmetric Ansatz \cite{qballs,bs,volkov}:
\begin{equation}
\psi=e^{i\omega t} \Phi(\theta) \ .
\end{equation}
Inserting the Ansatz into equation (\ref{cnls}), we find:
\begin{equation}
\label{equa}
\Phi''+\cot\theta\Phi' + \Phi'^2 \frac{g\Phi}{1+g\Phi^2} +
 \frac{4g\Phi^3-\omega \Phi}{1+g\Phi^2} R^2=0
\end{equation}
where the prime denotes the derivative with respect to $\theta$.
Note that we are fixing the length scale $\lambda$ of the space variable by
$R_{phy}=\lambda R$, where $R_{phy}$ is the physical radius of the
fullerene and $\lambda$ is of the order of \AA. We will vary $R$ here
in order to be able to take the $R\rightarrow\infty$ limit and thus compare 
our results with those of \cite{hz,bhz}. 
The other dimensionful physical quantities can be recovered by
reinserting the physical constants of the model appropriately (see e.g.
eq. (6), (7) of \cite{hz}). As far as the frequency $\omega_{phy}$ is concerned, we have:
\begin{equation}
\omega_{phy}=\frac{\hbar}{2m}\frac{1}{R_{phy}^2}\omega \ ,
\end{equation}
where $m$ is the mass of the ``electron''. Thus, since $\omega$ is dimensionless,
$\omega_{phy}$ has dimension of an inverse time: 
$[\omega_{phy}]=$sec$^{-1}$.

The regularity of the solutions at $\theta = 0,\pi$ requires the
following boundary conditions~:
\begin{equation}
\label{bc1}
\Phi'|_{\theta=0}=\Phi'|_{\theta=\pi}=0
\end{equation}
For later convenience, we denote by ${\cal M}$ the
reflexion operator with respect to the equatorial ($\theta=\pi/2$) plane~:
${\cal M}\Phi(\theta, \varphi) = \Phi(\pi-\theta, \varphi)$.
Solutions for which ${\cal M}\Phi=\Phi$ we will call symmetric, those
with ${\cal M}\Phi=-\Phi$ antisymmetric and finally solutions
for which ${\cal M}\Phi\neq\pm \Phi$ will be called asymmetric in the following.

It appears impossible to construct explicit solutions
of (\ref{equa}), (\ref{bc1}). This is why numerical techniques
have to be adopted. However, in our numerical study, it became obvious
that
the pattern of solutions occuring for different values of $\omega$ is
quite complicated. In an attempt to understand this pattern,
we have analysed the linearized equation first. This analysis
will be discussed prior to presenting the numerical results.
\subsection{Linearized equation}
First, we remark that (\ref{equa}) has three fixed points,
repectively
$\Phi_c = 0$ and $\Phi_c = \pm\sqrt{\frac{\omega}{4}}$. Perturbating
around
one of these fixed points according to
\begin{equation}
\label{peq}
\Phi(\theta) = \Phi_c + \epsilon f(\theta) + O(\epsilon^2)
\end{equation}
and linearizing the equation in $\epsilon$ leads to
\begin{equation}
f'' + \cot \theta f' - \omega R^2 f = 0 \ \  \ \ {\rm if} \ \
\Phi_c=0
\end{equation}
and
\begin{equation}
f'' + \cot \theta f' +
\frac{2 R^2 \omega}{1+\omega/4} f = 0 \ \  \ \ {\rm if} \ \
\Phi_c=\pm\sqrt{\frac{\omega}{4}}  \ .
\end{equation}
In both cases, we recognize the equation of Legendre polynomials $P_n$~:
$P_n'' + \cot \theta P_n' + n(n+1) P_n=0$.
If we take into account the boundary conditions given by (\ref{bc1}), the
 relevant solutions of the linearized equation are given 
in terms of these polynomials and fix a relation between $R$, $\omega$
and the integer $n$:
\begin{equation}
\label{p1}
f(\theta) = P_n(\theta) \ \ {\rm with} \ \ \omega R^2 = - n(n+1) \ \  \ \
{\rm if} \ \ \Phi_c = 0
\end{equation}
and
\begin{equation}
\label{p2}
\tilde f(\theta) = P_n(\theta) \ \ {\rm with} \ \
\frac{2\omega R^2}{1+\omega/4} =  n(n+1)  \ \  \ \
{\rm if} \ \ \Phi_c = \pm \sqrt{\frac{\omega}{4}} \ .
\end{equation}

In particular, the above relations
define two sets of  critical values of the spectral
parameter $\omega$ as a function
of $R$ which are indexed by $n$. For (\ref{p1}) and (\ref{p2}) we find respectively~:
\begin{equation}
\label{om1}
\omega_{cr,n}\equiv-\frac{n(n+1)}{R^2} \ ,
\end{equation}
\begin{equation}
\label{om2}
\tilde\omega_{cr,n} \equiv \frac{4n(n+1)}{8R^2-n(n+1)}  \ .
\end{equation}

From the above analysis, setting $n=1$ such that $P_1 \propto \cos \theta$,
we conclude that
the pertubated solution about
$\Phi_c = \sqrt{\omega/4}$ is  positive (i.e. especially  it
has no nodes, at least for values of $\omega$ close to the
critical values)
and has a maximum above the north pole ($\theta=0$), a minimum above
the south pole ($\theta=\pi$).

The perturbated solution about $\Phi_c = 0$ leads to  a solution which is
antisymmetric under the reflexion ${\cal M}$.
 It vanishes
at the equator (i.e. has a node at $\theta=\pi/2$) and
possesses a maximum above the north-pole, a minimum above the south pole.

Of course, the functions presented here constitute
only approximations (even up to an unfixed multiplicative constant) 
to the solutions of the full equation (\ref{equa}).
The closer we are to the critical points (\ref{om1})-(\ref{om2}), 
the more accurate
these approximations are.  Nevertheless, preforming this analyis we gain
extremely useful information about the solutions of
the full equations, namely about
(i) their symmetries and (ii) the critical values of $\omega$
where the various branches start or finish.
In the following, we will discuss the numerical results.

\subsection{Numerical results}
We constructed solutions numerically for generic values of $R$ and
of $\omega$. In our numerical routine, $\Phi(0)$ is used as a shooting
parameter and the regular solution as well as the 
corresponding value of $\omega$ are determined numerically.
We first constructed a ``fundamental" solution which has no
node ($k=0$). For generic values of the parameters, this solution
corresponds to the above mentioned deformation of the Legendre polynomial
$P_1$ and is thus
positive, has a maximum
for $\theta=0$ and a minimum at $\theta= \pi$.
This is illustrated in Fig.1 for $\Phi(0)=1.0$, $0.5$ and $0.3$.
Keeping $R$ fixed  and varying $\Phi(0)$
we obtain a branch of asymmetric solutions labelled by $\omega$. Our
numerical analysis strongly suggests that these regular
solutions exist for $ \tilde\omega_{cr,1} \leq \omega < \infty$ with
\begin{equation}
       \tilde\omega_{cr,1} \equiv \frac{1}{R^2 - 1/4}  \ .
\end{equation}
In the limit $\omega \rightarrow \tilde\omega_{cr,1}$ the solution tends 
to the
constant solution $\Phi = \sqrt{\tilde\omega_{cr,1}/4}$, which for $R=2$ 
is equal
to 0.2582. As can be seen in Fig.1, the solution for $\Phi(0)=0.3$
is already nearly constant and thus close to this critical solution.

In correspondence with these solutions, the mirror symmetric
ones  also exist and constitute
another branch with the same values of $\omega$.

Next, we  constructed a family of solutions having one node
for $\theta \in ]0, \pi[$.
The node occurs at
 $\theta = \pi/2$ and the  solutions
 of this branch are antisymmetric under the
reflection  ${\cal M}$, as illustrated in Fig. 2.
These solutions exist only for $\omega \geq 
\omega_{cr,1} \equiv -2/R^2$
and for arbitrary large values of $\omega$.
In the limit
$\omega \rightarrow \omega_{cr,1}$, the solutions are of the
form (\ref{peq}) with $\Phi_c=0$ and $f(\theta)=\cos\theta$. This is
shown in Fig.2, where the solution with $\Phi(0)=0.25$ is proportional to
the trigonometric function $\cos\theta$.
 When $\omega$ is increased
the effects of the non-linear term become stronger
and the solution's profile
deviates considerably from an elementary  trigonometric function.

When investigating this branch of solutions for larger values
of $\Phi(0)$ (or equivalently of $\omega$) we observed that
the one-node solution develops a plateau (surrounding the region
$\theta = \pi/2$) on which the function $\Phi$ is very close to zero.
Away from this plateau the solution resembles two 
disconnected solitons located at the two poles of the sphere. 
This is illustrated in Fig. 3 (solid line) for
$\Phi(0) = 3.7$ and the corresponding value of $\omega$ is
$\omega =15.95$. However, for large enough values 
of $\Phi(0)$, i.e. of $\omega$, our 
numerical analysis strongly indicates that {\it asymmetric} 
solutions exist next to the antisymmetric ones
described above.  
In Fig. 3 such a asymmetric solution corresponding to $\omega= 14.85$
together with 
a comparable antisymmetric solution is presented.
The asymmetric solution in Fif. 3 (dotted line)
has an energy of order -900 (in our dimensionless variables). 
This has to be contrasted with the energy of the antisymmetric
solution (solid line) which is of order -300.
This result is interesting with view to \cite{harigaya}. There, it was found
that the excess electron density in the C$_{60}$ fullerene
is located along an equatorial line of the C$_{60}$. The excess
electron density here can be represented by $\Phi^2(\theta)$. We find here,
that the solution with the electron density located at the equator (asymmetric
solution) is energetically more favourable than the one which has the
electron density located in one of the hemispheres.
 It is remarkable that the two solutions
show a very similar behaviour around $\theta = 0$ and start
to deviate from each other at roughly $\theta > 0.8$. 
Unfortunately, this similarity 
of the solutions in the region around the north pole renders the
numerical 
construction of the solutions extremely difficult. This is why
we did not succeed to construct the full asymmetric branch.
However, let us mention that
a similar feature occurs as well in the spinning case $N = 1$. In this case,
the numerical analysis was much easier and we will thus discuss this point
again in the section about spinning solutions.

The values of the norm $\eta$, the energy $E$ and the value $\Phi(0)$, which
 corresponds to the maximal value of the solution,
 are shown in Fig. 4
as functions of $\omega$ for both the fundamental ($k=0$) and the one-node
($k=1$) solution with $R=2$ (the graphic is limited to $\omega \leq 1.5$).

Solutions having more nodes of the function
$\Phi(\theta)$ can be constructed in correspondence with the
higher order Legendre polynomials. However, the systematic study
of these solutions is not the aim of this paper.
 
 Finally, we remark that
\begin{equation}
\lim_{R \rightarrow \infty} \omega_{cr,1} = \lim_{R \rightarrow \infty}
\tilde\omega_{cr,1} = 0   \ .
\end{equation}
This is  in agreement with \cite{bhz},
and confirms that in the plane
both fundamental and one-node
solutions can be constructed for $\omega \geq 0$.

\section{Spinning solutions}
In order to construct spinning solitons,
we use the  Ansatz \cite{bs,volkov}:
\begin{equation}
\psi=e^{i\omega t + i N\varphi}\Phi(\theta)  \ .
\end{equation}
which, after inserting  into (\ref{cnls}), leads to the equation :
\begin{equation}
\Phi''+\cot \theta \Phi' +
\Phi'^2\frac{g\Phi}{1+g\Phi^2}+\frac{4g\Phi^3-\omega\Phi}{1+g\Phi^2}R^2 
- \frac{N^2}{1+g\Phi^2} \frac{1}{\sin^2 \theta } \Phi=0  \ .
\end{equation}
 Regularity of the solutions at $\theta = 0, \pi$ requires
 \begin{equation}
 \Phi(0) = \Phi( \pi) = 0   \ .
 \end{equation}
 In this case, $\Phi_c = 0$ is the only fixed point of the equation
 and the corresponding linearized equation is solved by the
 associated Legendre functions $P_n^N$ , $-n \leq N \leq n$,
 provided the relation (\ref{om1}) holds.
In what follows, $n$ will be related to the number $k$ of nodes of
the solution by $k=n-N$.
 
 We would like to mention that we were able to construct an exact
 solution of the above equation for $N=1$ and $R^2 = 3/4$. It
 is of the form 
 \begin{equation}
 \label{asol}
 \Phi = \alpha \sin \theta \ \ \ {\rm with} \ \ \ \omega=\frac{8(\alpha^2-1)}{3} \ ,
 \end{equation}
 where $\alpha$ is a real
 constant. 
Obviously, this solution is symmetric.

\subsection{Numerical results for $N=1$}
In this case, the numerical solutions can be constructed by
using the value $\Phi'(0)$ as a shooting parameter.
We studied numerically the solutions corresponding to the
deformation of the associated Legendre function
$P_1^1 \propto \sin \theta $ (for no-node solutions) and of
$P_2^1 \propto \sin \theta \cos \theta $ (for one-node solutions) and 
found that 
in accordance with (\ref{om1}), there exist no solutions for respectively
\begin{equation}
      \omega \leq -\frac{2}{R^2}\equiv \omega_{cr,1}  
      \ \ , \ \ \omega \leq -\frac{6}{R^2}\equiv \omega_{cr,2}
\end{equation}
As for the case of non-spinning solution, the norm $\eta$, the energy
$E$ and the parameter $\Phi'(0)$ are presented in Fig. 5.

As can be seen from this figure, a new phenomenon occurs in the case
of spinning solutions without nodes. Namely, the main 
branch corresponding to
the deformation of $P_1^1$ stops at some critical value
of $\omega$, say $\omega_{max}$ with $\omega_{max} > \omega_{cr,1}$ 
(for $R=2$, we find $\omega_{max} \approx 0.3$).
However, the inspection of Fig. 5 by no means reveals that some critical
phenomenon occurs in this limit.
When plotting the  quantities $E$, $\omega$ and the
norm $\eta$ as functions of the
shooting parameter $\Phi'(0)$  (see Fig. 6), though, it is clearly indicated that
some critical phenomenon appears. Indeed   as our numerical
results indicate, the slope of the curves becomes infinite
when the shooting parameter reaches the critical point.
The solutions on this branch  are symmetric under
the operator ${\cal M}$. The profile of the symmetric solution
corresponding to $\Phi'(0) = 0.25$ is shown in Fig.7.

For sufficiently large values of $\omega$,
say $\omega > \overline\omega$, (
$\overline\omega \approx 1.0$ for $R=2$)
a new branch of nodeless solutions
appears which seems to exist for arbitrary large values of $\omega$, while
for $\overline\omega > \omega >\omega_{max}$ no solutions seem to exist at all.

The solutions for $\omega > \overline\omega$ are asymmetric under the 
reflexion
${\cal M}$ . Only the limiting solution at $\omega=\overline\omega$ 
is symmetric under ${\cal M}$.
An asymmetric solution corresponding to $\Phi'(0)=1.0$ as well as the limiting,
symmetric solution for  $\Phi'(0)= 0.36$ is shown in Fig.7. Clearly the 
asymmetry increases with increasing $\Phi'(0)$.
In other words, when $\omega >\overline \omega$ increases, the wave function
of spinning, nodeless solitons is more and more concentrated in one of the
hemispheres.

All the above results were obtained for $R=2$. However, we also studied
the dependence of the critical values of $\omega$ on the radius $R$ of the sphere. The results
are given in Table 1.\\
\\
\\
\begin{center}
Table 1: Critical values of $\omega$ for different $R$\\
\medskip
\begin{tabular}{|c|c|c|c|}
\hline
$R$ & $\omega_{cr,1}$ & $\omega_{max}$ & $\overline\omega$ \\
\hline \hline
$\infty$ & $0$ & $0$ & $0$ \\
\hline
$5$ & $-2/25$ & $0.03$ & $0.15$\\
\hline
$2$ & $-1/2$ & $0.3$ & $1.0$ \\
\hline
$1.5$ & $-2/(1.5)^2$ & $1.3$ & $2.15$ \\
\hline
$\sqrt{3/4}$ & $-3/8$ & $\infty$ & $\infty$\\
\hline
\end{tabular}
\end{center}

For generic value of  $R$  such that $\sqrt{3/4} < R < \infty$ two branches of solutions
exist, namely a branch of symmetric solutions for $\omega \in 
[\omega_{cr,1}:\omega_{max}]$
and  a branch of asymmetric solutions for $\omega \in 
[\overline\omega:\infty]$. 
In the limit $R\rightarrow\infty$ , the symmetric branch disappears and only the asymmetric one
survives. In the limit $R\rightarrow \sqrt{3/4}$ only symmetric solutions seem to exist
since $\overline\omega \rightarrow \infty$. In this respect, the explicit solution
(\ref{asol}) plays a major role.

Finally, we constructed one branch of spinning, one-node solutions.
They exist for $\omega \geq -6/R^2\equiv \omega_{cr,2}$  and are symmetric under
the operator ${\cal M}$. They vanish when $\omega$ approaches
the lower critical limit $\omega_{cr,2}$ and seem to exist for arbitrary large values
of $\omega$. Spinning, one node solutions corresponding to $\Phi'(0)=1.0$, $2.0$ and $3.0$ are shown in
Fig. 8.\\
\\
\subsection{Numerical results for $N=2$}
We have also constructed spinning solutions corresponding to $N=2$. The solutions
are deformations of the associated Legendre functions $P_2^2\propto 1-\cos(2\theta)$ (for no-node
solutions), respectively $P_3^2\propto \cos\theta-\cos(3\theta)$ (for one-node solutions).
In agreement with the analyis of the linearized equation, no solutions exist respectively
for 
\begin{equation}
\omega \leq -\frac{6}{R^2}\equiv\omega_{cr,2} \ \ , 
\ \ \omega \leq -\frac{12}{R^2}\equiv\omega_{cr,3} \ .
\end{equation}
Furthermore, the no-node solutions are symmetric under ${\cal M}$, while the
one-node solutions are antisymmetric under ${\cal M}$. No branch of asymmetric solutions seems
to exist in this case.
Thus, as far as our numerical analysis indicates, in contrast to the $N=1$ case,
 these are the only
$N=2$ solutions for $n \leq 3$. 

\section{Conclusion}
The study of the modified non-linear Schr\"odinger equation of
\cite{hz,bhz,bepz} on the sphere
reveals a very rich set of stationary solutions. They can be characterised by their behaviour
under the reflection operator ${\cal M}$ with respect to the equator ($\theta=\pi/2$) plane.
We constructed branches of solutions which are either
 i) symmetric under ${\cal M}$, ii) antisymmetric
under ${\cal M}$ or iii) asymmetric under ${\cal M}$. The solutions are further characterised
by their spinning number $N$ and the number of nodes $k$ of the wave function
which we find -as expected- to be closely related to the quantum numbers of the spherical
harmonics. Linearizing the modified non-linear Schr\"odinger equation, we find that
the lower value of $\omega$, above which solutions exist, is determined
by the spectral parameters obtained from the linearized equation.

An unexpected phenomenon occurs in the cases $N=0$, $k=1$ and $N=1$, $k=0$. 
For $N=0$, $k=1$, we find that for comparable values of $\omega$
two branches of solutions exist: one which is antisymmetric and
one which is asymmetric.
For $N=1$, $k=0$, we observe that for finite $R> \sqrt{3/4}$ the symmetric
solutions only exist on a finite interval of the parameter $\omega$.  
Since $\omega$ is related to the kinetic energy of the solution \cite{bhz},
this means in other words that only for solutions with sufficiently low kinetic energy
is the probability density of the ``electron''  the same in both hemispheres. If the kinetic
energy becomes large enough, the electron density is essentially concentrated
in one of the hemispheres.
In \cite{harigaya}, it was found that the excess electron density in fullerides
is located around the equator. Here, we find $\Phi^2(\theta)$ to be concentrated
around the equator in two cases: (a) for the $N=0$, $k=1$ case if $\omega$
is large enough, (b) for $N=1$, $k=0$ if $\omega\in [\omega_{cr,1}(R):\omega_{max}(R)]$.

We have not studied $N > 2$ and/or $k>1$ in this paper. We believe that the above mentioned
properties hold also true for these type of solutions and that, in particular, the
solution with $N$, $k$ is related to the associated 
Legendre function $P^N_{k+N}$.\\
\\
\\
{\bf Acknowledgments} Y.B. gratefully acknowledges the Belgian F.N.R.S.
for financial support. B.H. gratefully acknowledges discussions with W. Zakrzewski.

 \newpage
\begin{figure}
\centering
\epsfysize=20cm
\mbox{\epsffile{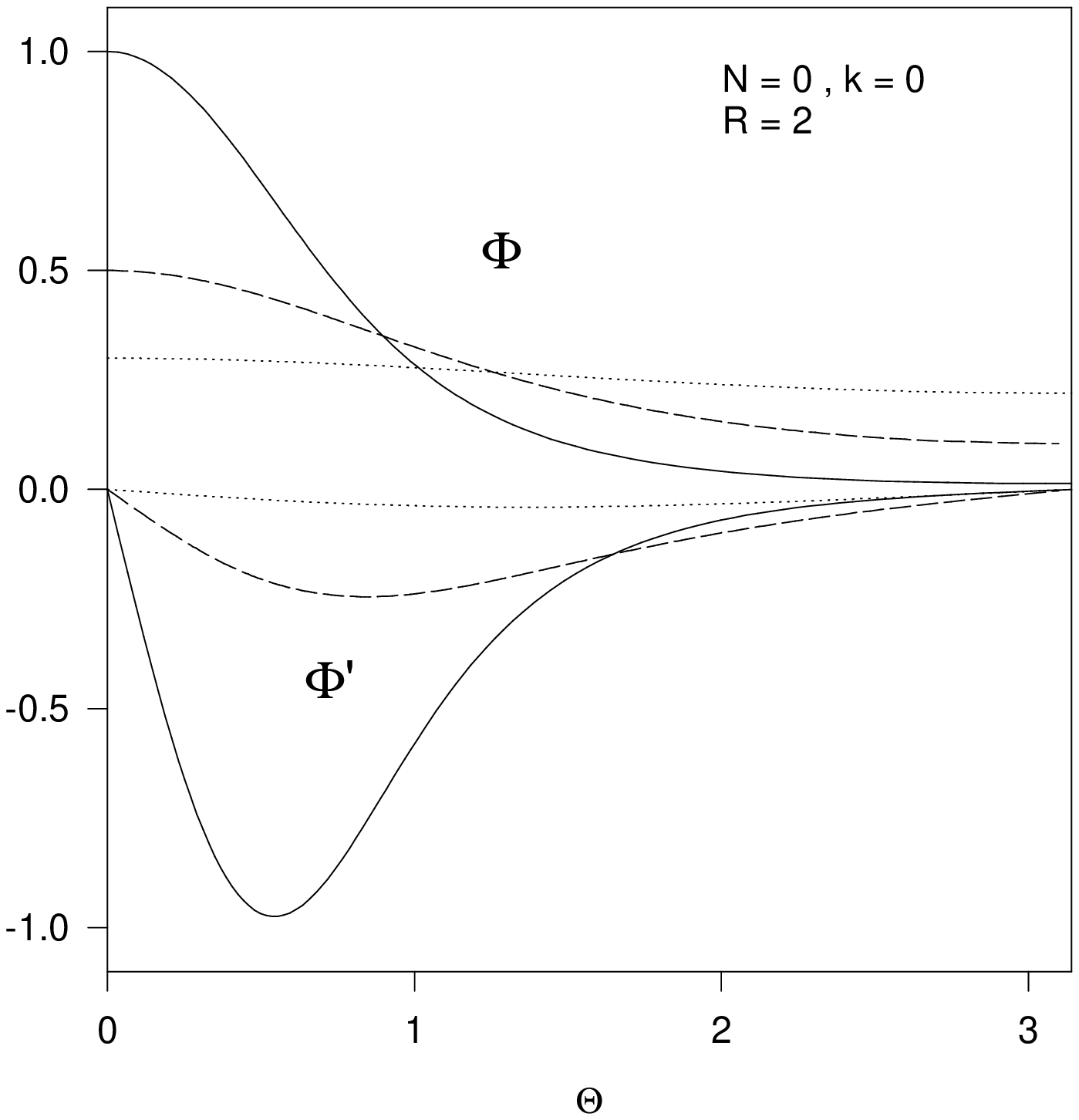}}
\caption{\label{Fig.1} The profile of the non-spinning, nodeless 
($N=0$, $k=0$) solution 
and its derivative is presented
for $R=2$ and three values of $\Phi(0)$, namely $\Phi(0)=1.0$, $0.5$ and $0.3$.}
\end{figure}
 \newpage
\begin{figure}
\centering
\epsfysize=20cm
\mbox{\epsffile{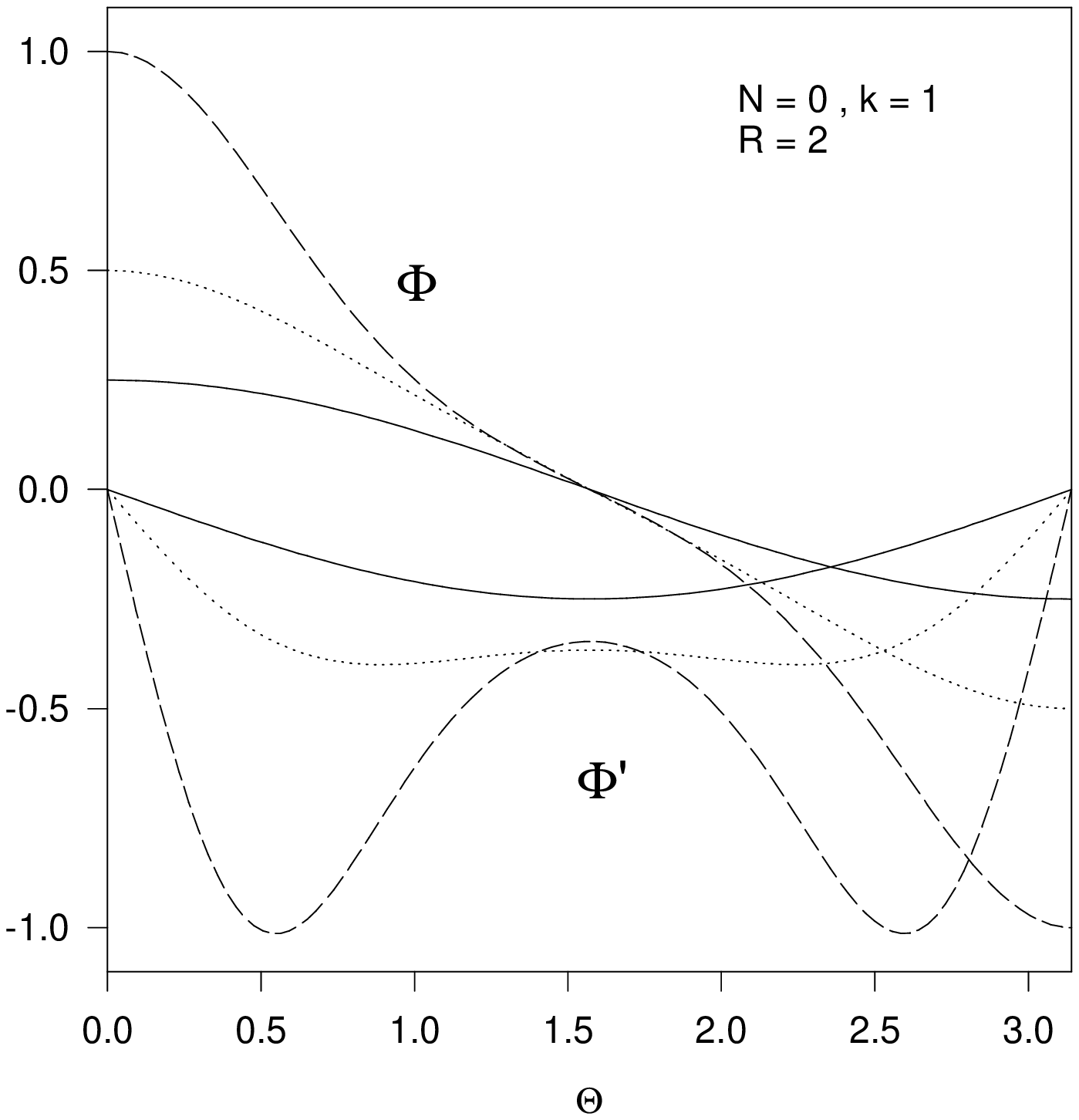}}
\caption{\label{Fig.2}The profile of the non-spinning, one-node 
($N=0$, $k=1$) solution
and its derivative is presented
for $R=2$ and three values of $\Phi(0)$, namely $\Phi(0)=1.0$, $0.5$ and $0.25$.}
\end{figure}
 \newpage
\begin{figure}
\centering
\epsfysize=20cm
\mbox{\epsffile{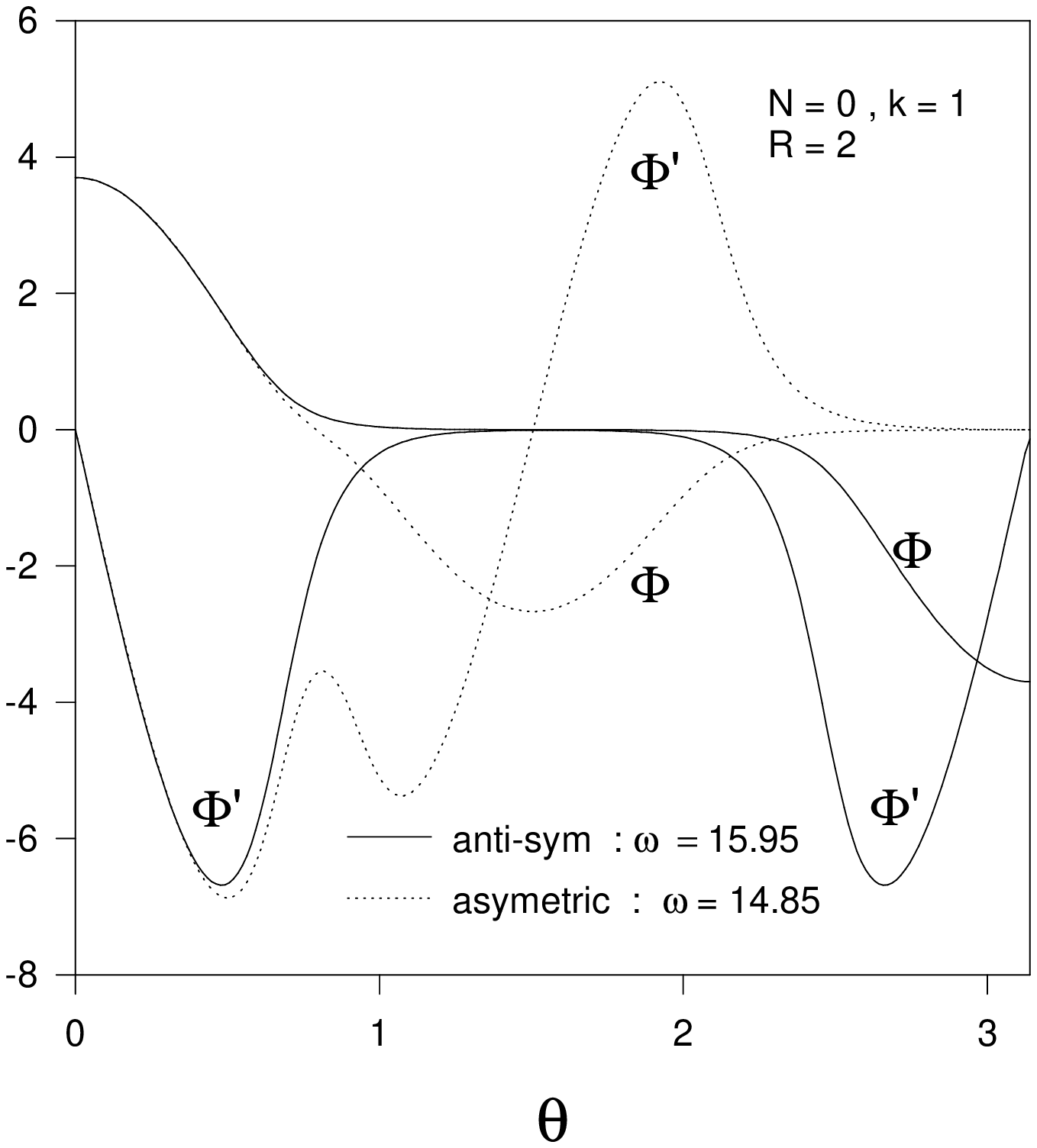}}
\caption{\label{Fig.3}The profile of 
one solution on the antisymmetric branch (corresponding to
$\omega=15.95$) as well as one solution on the asymmetric  
branch (corresponding to $\omega=14.85$) is shown for $N=0$, $k=1$ and $R=2$.
The corresponding derivatives of the solutions are also shown.}
\end{figure}
 \newpage
\begin{figure}
\centering
\epsfysize=20cm
\mbox{\epsffile{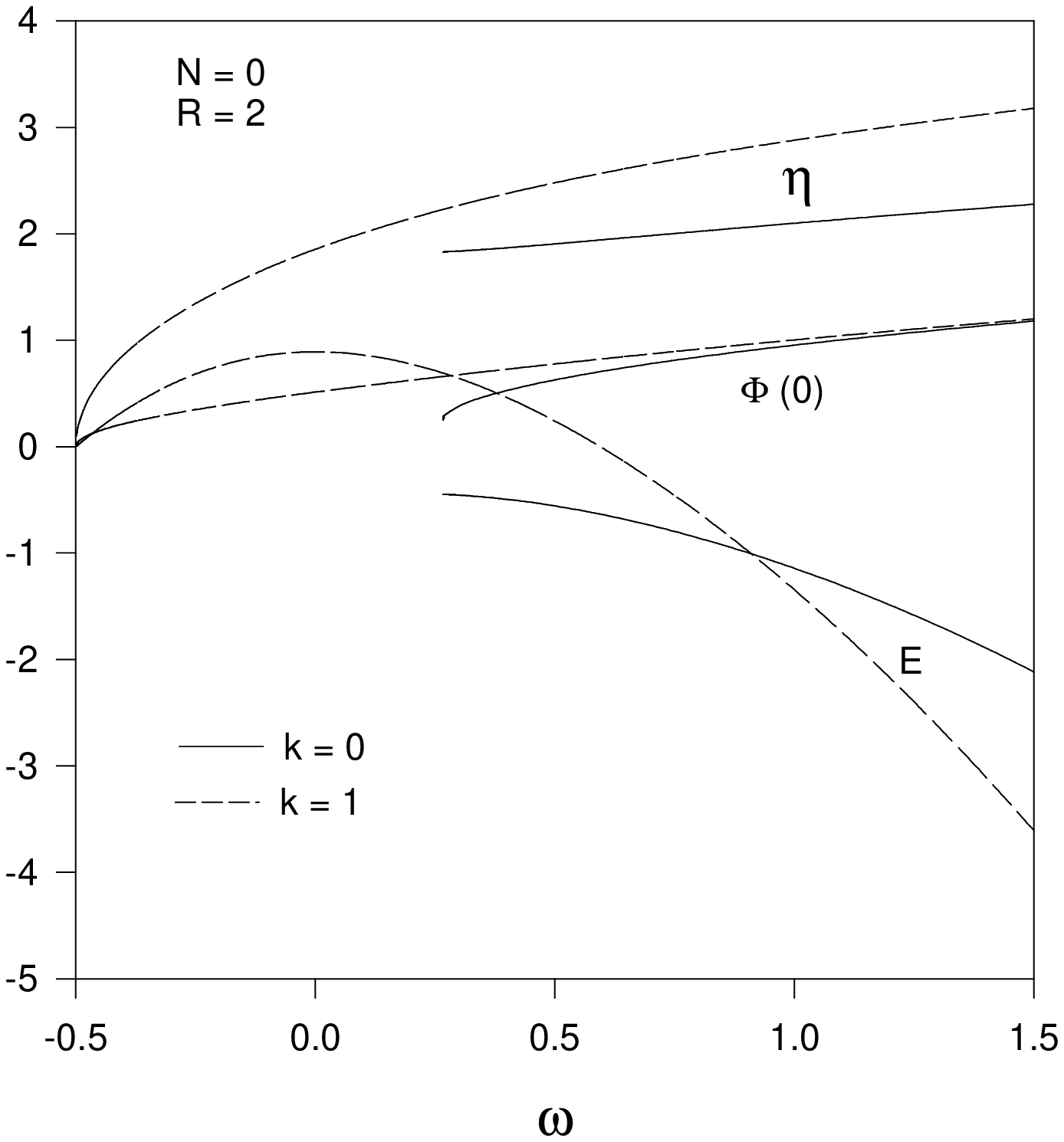}}
\caption{\label{Fig.4}The values of the energy $E$, the norm $\eta$ and of $\Phi(0)$
are shown as functions of $\omega$ for the fundamental ($k=0$)
and one-node ($k=1$) solutions with $N=0$ (non-spinning) and $R=2$.}
\end{figure}
 \newpage
\begin{figure}
\centering
\epsfysize=20cm
\mbox{\epsffile{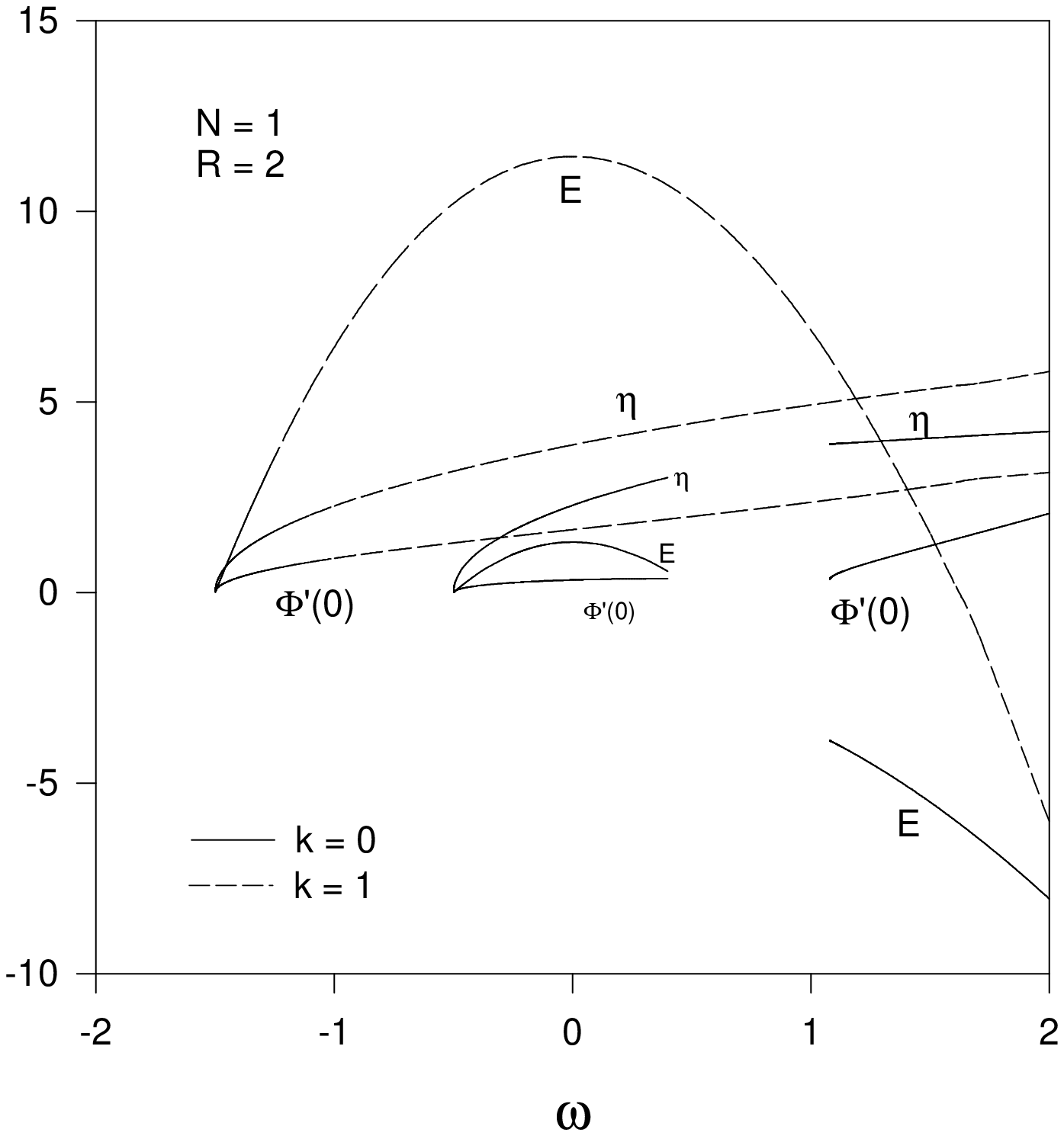}}
\caption{\label{Fig.5}The values of the energy $E$, the norm $\eta$ and of $\Phi'(0)$
are shown as functions of $\omega$ for the fundamental ($k=0$)
and one-node ($k=1$) solutions with $N=1$ (spinning) and $R=2$.}
\end{figure}
 \newpage
\begin{figure}
\centering
\epsfysize=20cm
\mbox{\epsffile{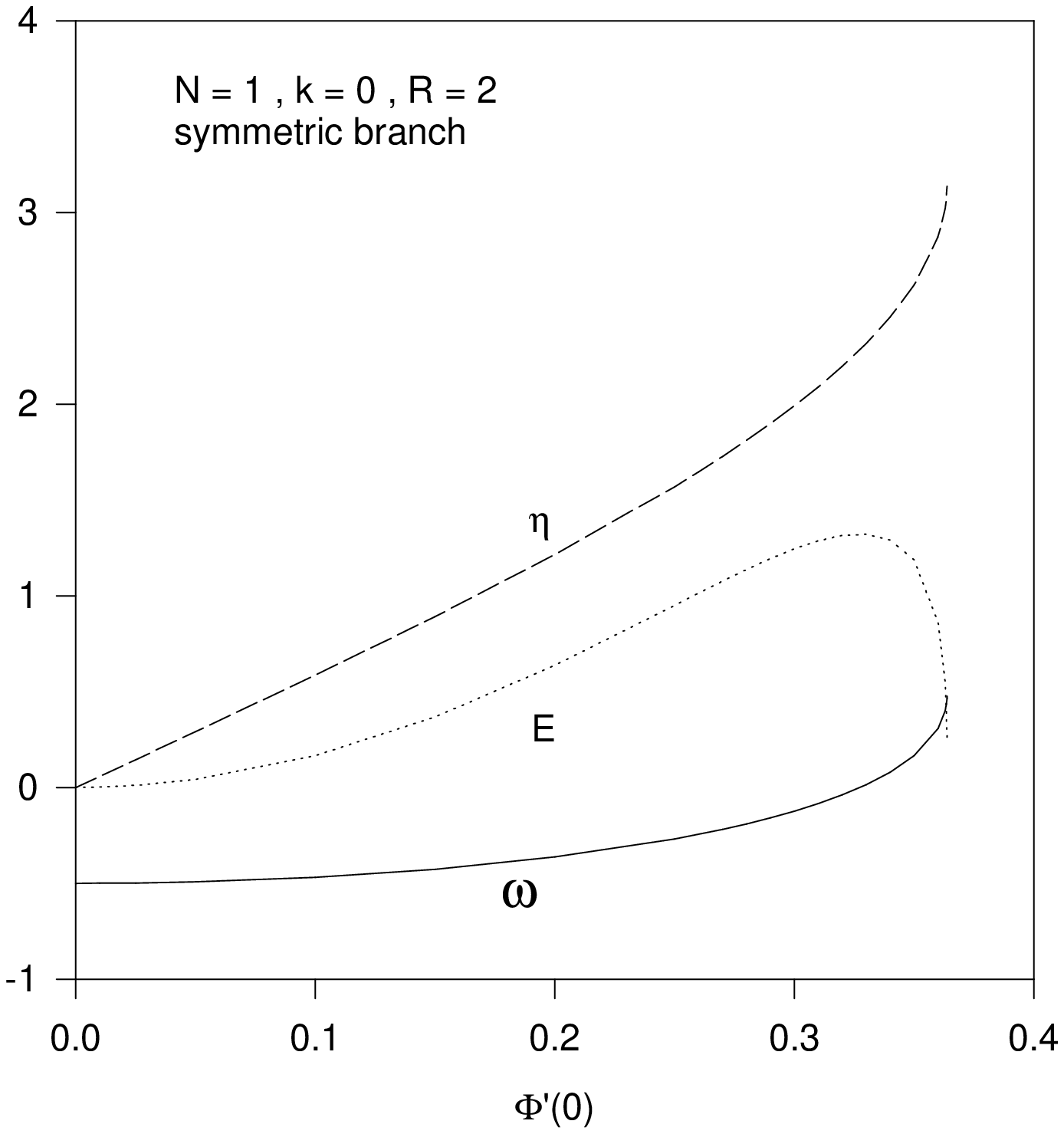}}
\caption{\label{Fig.6}The values of the energy $E$, the norm $\eta$ and of $\omega$
are shown as function of $\Phi'(0)$ for the spinning, nodeless ($N=1$, $k=0$)
solutions with $R=2$.}
\end{figure}
 \newpage
\begin{figure}
\centering
\epsfysize=20cm
\mbox{\epsffile{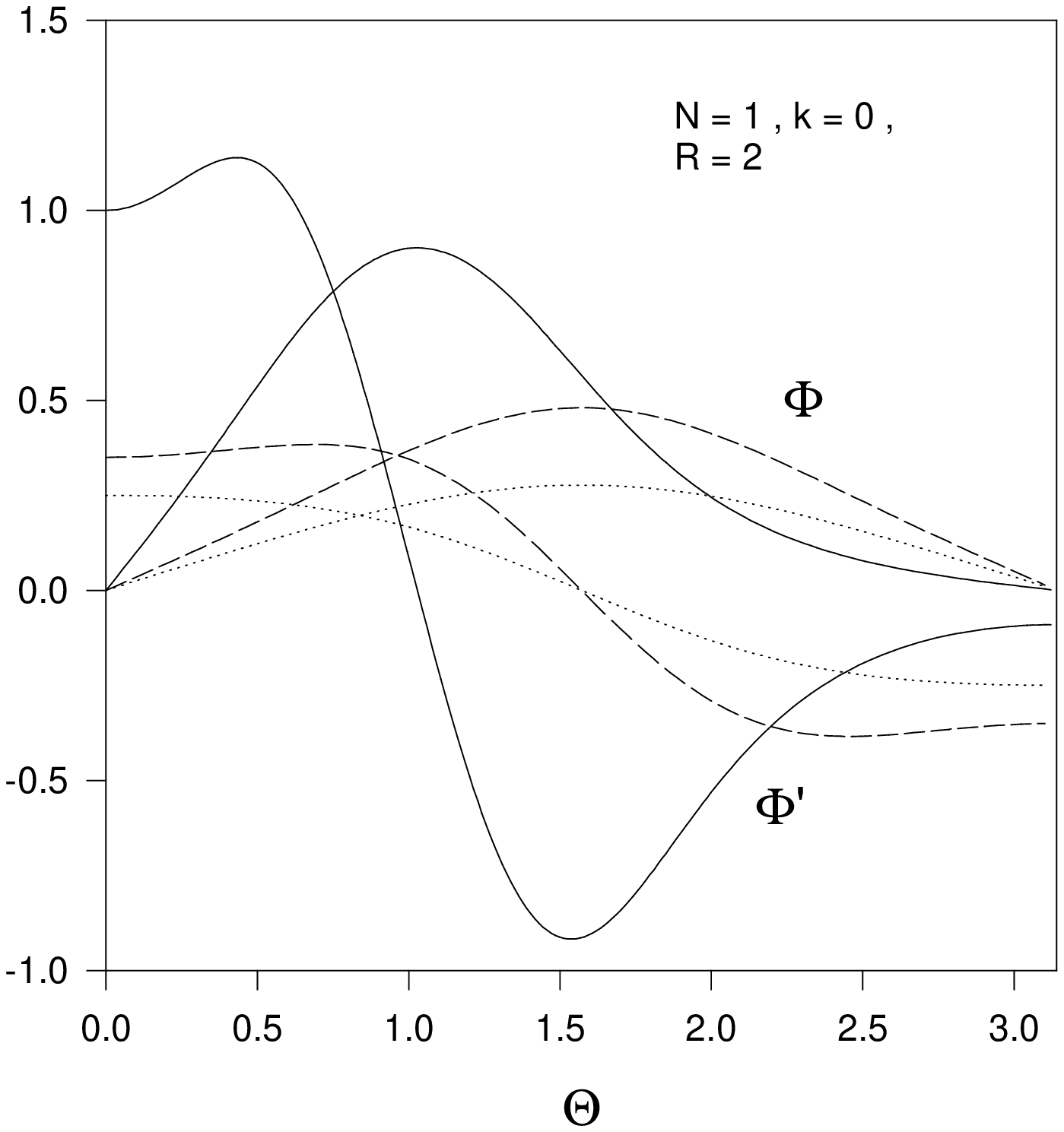}}
\caption{\label{Fig.7}The profile of the spinning, nodeless 
($N=1$, $k=0$) solution
and its derivative is shown
for $R=2$ and three values of $\Phi'(0)$, namely $\Phi'(0)=1.0$, $0.36$ and
$0.25$.}
\end{figure}
\begin{figure}
\centering
\epsfysize=20cm
\mbox{\epsffile{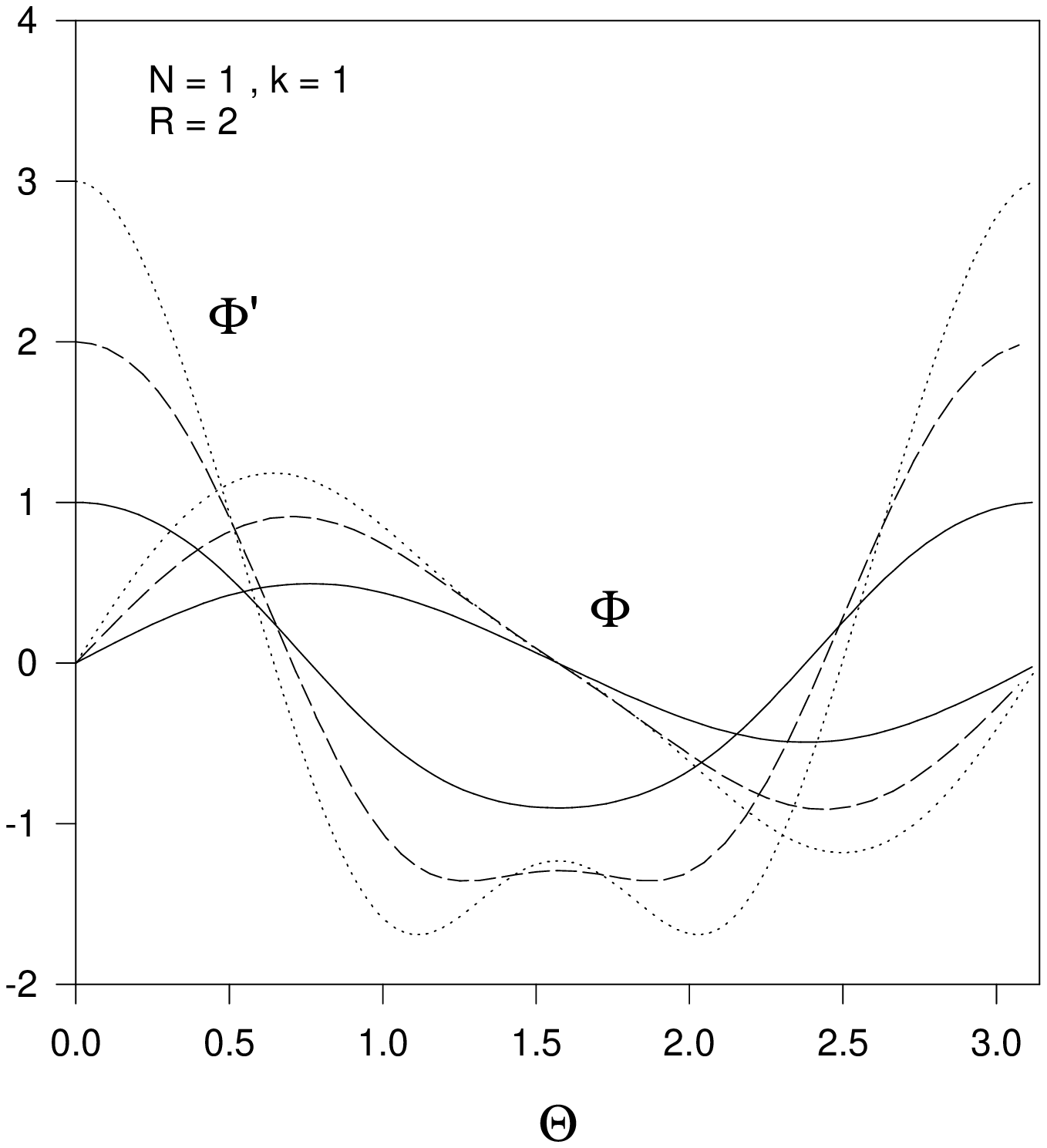}}
\caption{\label{Fig.8}The profile of the spinning, one-node ($N=1$, $k=1$) solution
and its derivative is shown
for $R=2$ and three values of $\Phi'(0)$, namely $\Phi'(0)=3.0$, $2.0$ and
$1.0$.}
\end{figure}
 \end{document}